# HYBRID CHAOTIC METHOD FOR MEDICAL IMAGES CIPHERING

Seham Muawadh Ali Ebrahim

Department of Computer Engineering and Information Technology, Modern Academy for Engineering and Information Technology, Egypt

*ABSTRACT*

*Healthcare is an essential application of e-services, where for diagnostic testing, medical imaging acquiring, processing, analysis, storage, and protection are used. Image ciphering during storage and transmission over the networks used has seen implemented using many types of ciphering algorithms for security purpose. Current cyphering algorithms are classified into two types: traditional classical cryptography using standard algorithms (DES, AES, IDEA, RC5, RSA, ...) and chaos cryptography using continuous (Chau, Rossler, Lorenz, ...) or discreet (Logistics, Henon, ...) algorithms. The traditional algorithms have struggled to combat image data as compared to regular textual data. Whereas, the chaotic algorithms are more efficient for image ciphering. The Significancecharacteristics of chaos are its extreme sensitivity to initial conditions and algorithm parameters.*

*In this paper, medical image security based on hybrid/mixed chaotic algorithms is proposed. The proposed method is implemented using MATLAB. Where the image of the Retina of the Eye to detect Blood Vessels is ciphered. The Pseudo-Random Numbers Generators (PRNGs) from the different chaotic algorithms are implemented, and their statistical properties are evaluated using the National Institute of Standards and Technology NIST and other statistical test-suits. Then, these algorithms are used to secure the data, where the statistical properties of the cipher-text are also tested. We propose two PRNGs to increase the complexity of the PRNGs and to allow many of the NIST statistical tests to be passed: one based on two-hybrid mixed chaotic logistic maps and one based on two-hybrid mixed chaotic Henon maps, where each chaotic algorithm runs side-by-side andstarts with random initial conditions and parameters (encryption keys). The resulting hybrid PRNGs passed many of the NIST statistical test suits.*

## 1. INTRODUCTION

Images security needs attention from the organizations that implement their applications.

Military, government, civilian, and industrial organizations are transmitting, receiving, and storing a large sized of datafor decision-making, security, and for other types of applications including healthcare, video surveillance, and Internet of Things (IoT) systems [3][4].

Images security requirements arise from the need to protect its data: first, from accidental loss and corruption, and second, from deliberate unauthorized attempts to access or alter that data. Cryptography is generally considered as the core and the best method of Images protection against passive and active frauds.

A cryptographic system (cipher or cryptosystem) converts Images data, referred to as "plaintext" into an encrypted format, referred to as "cipher-text" to be stored or send over an unsecured channel. The encryption process is accomplished by ciphering, manipulating or transforming the plaintext into the cipher-text by the ciphering algorithm using a "cypher key" or keys.

The receiver "decrypts" the cipher-text, that is, converts it from cipher-text to plaintext, by reversing the manipulation or transformation process using the decipher key or keys.

When only the sender and receiver have access to the cipher and decipher algorithms and keys, such an encrypted transmission is safe.

A "traditional" symmetrical cryptosystem (e.g. DES, 3DES, IDEA, AES, and RC5) is a cryptosystem in which it is possible to use the enciphering data to evaluate the deciphering data. This cryptosystem requires the encryption key to be kept secret and given to the user's device through secure channels such as secret couriers, secure telephone transmission lines, or the like, in order to provide protection.

Asymmetrical or public-key encryption (e.g. RSA, and El Gamal) is regarded as a method that avoids the difficulties of exchanging a protected encryption key. An enciphering function is chosen so that once an enciphering key is known; the enciphering function is relatively easy to compute. However, the inverse of the encrypting transformation function is difficult, or computationally infeasible, to compute. Such a function is called "one-way function" or "trap door function." [5]

The most common encryption objects are block-encryption algorithms (secret-key algorithms or symmetric algorithms), PRNGs (additive stream ciphers), and public-key algorithms (asymmetric algorithms). Block ciphers transform a relatively short string (typically 64, 128, or 256 bits) to a string of the same length under the control of a secret key.

Whereas traditional ciphering algorithms are used efficiently for text or binary data, they are not ideal for Images, video, and audio data, which are characterized by the massive size of data and require high bit rate and bandwidth (near real-time communication). [9].

Chaotic encryption is a direction of cryptography. It makes use of chaotic system properties such as sensitivity to initial conditions and dependence on the system parameters. Many chaos-based encryption methods are presented and discussed [7] [8] [9].

A PRNG is a deterministic method to generate from a small set of random numbers, called the seed, a more extensive set of pseudo-random numbers. Chaotic systems are efficiently used as PRNGs. Statistical properties of binary sequences generated by Chaotic-based PRNGs can derive a sufficient condition for the class of chaotic-maps to produce a sequence of independent and identically distributed binary random variables are efficiently used. Images security by cryptography requires the use of PRNGs, which is considered in this paper to be implemented by chaotic systems since their outputs are sequences of statistically independent and unbiased numbers. Chaos related researches are focused on the applications of chaotic algorithms for securing all types of data, digital identity, biometrics identifications, and communications.

This paper simulates and tests the performance of chaotic encryption when used to secure Images. Various continuous and discreet chaotic-algorithms are implemented using Matlab simulations are used to secure Images video signals. The PRNGs of the algorithms are studied, and their statistical performances are evaluated using NIST [12] and other test-suites[13]. To improve the statistical properties of the Chaotic-algorithms, we implemented two-hybrid Chaotic-algorithms consisting of two Logistics maps and two Henon maps running side by side and having different parameters and initial conditions. Compared with single Chaotic-algorithms, it was proved that the statistical properties of hybrid mixed algorithms passed most of the NIST test-suits and, therefore, can be used more effectively to Images security.

## 2. LITERATURE REVIEW

In [1] an Image Encryption and Decryption Based on Chaotic Algorithm is presented. This paper introduces the use of the two-stage logistic algorithm in image encryption and decryption and verifies via more secure information entropy. The paper proposes a new algorithm for image encryption and decryption based on the combination of multiple sequences and two-stage logistic maps.

In [2] a new way of image encryption scheme using the development of chaotic logistic map based on feedback stream cipher using an external secret key of 256-bit. Logistic maps are developed to increase the range of its variables. Several test images are used for inspecting the validity of the proposed algorithm. The robustness of the proposed algorithm is based on a feedback mechanism that leads the cipher to a cyclic behaviour so that every single pixel's encryption depends on the performance of the chaotic map used and the previous cipher pixel. Also [6] present an image encryption using chaotic map which is based on 1D logistic maps.

In [10], a description of various video encryption algorithms based on traditional and chaotic algorithms are presented. The analysis is concerning some parameters like encryption speed, security level, and stream size. It was concluded that it is difficult for a particular algorithm to satisfy all security requirements parameters. Therefore, the encryption algorithm can be chosen based on the application specifications in use.

In [11], the asymmetric encryption algorithm based on bit permutations and using an iterative process combined with a chaotic function. The main advantages of such a cipher system are its ability to encrypt bit sequences securely and assuring confusion, diffusion, and indistinguishability properties in the cipher.

In [13], an algorithm for the image encryption/decryption scheme using multiple chaotic based circular mapping is presented. First, pairs of subkeys are given in the paper using chaotic logistic maps. Second, using the logistic map subkey, the image is encrypted and contributes to the process of diffusion in its transformation. Third, subkeys are generated by four separate chaotic maps. Each map will generate different random numbers from the different orbits of the maps, based on the initial conditions. also, in [16]proposed block cipherbased on multiple chaotic systemsto resist three weakness: The Chosen Plaintext AttacksCPAs, the Chosen Ciphertext AttacksCCAsand the Known Plaintext AttacksKPAs.

## 3. BLOCK DIAGRAM OF CHAOTIC-BASED IMAGES SECURITY

The general block diagram of the simple symmetric chaotic ciphering system is shown in figure 1.

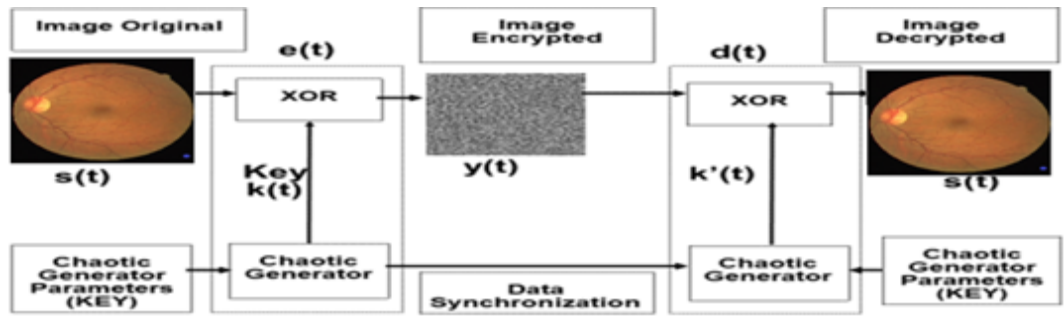

Figure 1: The general block diagram of the simple symmetric chaotic cypher system.

The system is explained as follows. The chaotic cipher system consists of encrypted and decrypted parts.

The encryption part *e(t)* consists of a chaotic system model (controlled by the encryption key signal) and an encryption algorithm. *K(t)* is the key signal, one of the state variables of the chaotic system, and is applied on *e(t)* to the image original state variable $s(t)$ to produce the image Encrypted transmitted signal *y (t)*.

The decryption part consists of a chaotic system and a decryption function *d(t)* and a decryption key $\dot{k}(t)$ which can equal *k(t)* or related to it via the data synchronization. The decryption can find the key signal when the decrypted and the encrypted are synchronized.

The encrypted signal can be recovered via synchronization, and then $d(t)$ is used to decrypt the encrypted signal.

This system represents the continuous system, where it differs from the traditional discrete cipher system as the key and the encrypted signal should be transmitted from the encrypted part to the decrypted part.

## 4. IMAGES SECURITY USING CHAOTIC SIGNAL GENERATIONS

This section discusses the implementation of different continuous and discrete-time chaotic generators in image security [5][14]. Each Images security scheme is discussed via three aspects:

1) The Matlab Simulation of the chaotic algorithms using Matlab (.m and Simulink) codes.
2) The application of the chaotic algorithm to secure generic Image files.
3) The statistical tests of the cipher-text where the output of each scheme shall be demonstrated and tested using NIST [12] and other PRNG test batteries.

### 4.1. Chaotic-based PRNG

#### 4.1.1. Continuous Chaotic Generators:

a. **Chua chaotic generators**

Chua's is a simple dynamic chaotic algorithm that exhibits chaos behavior under certain initial conditions and the following differential equations parameters:

International Journal of Network Security & Its Applications (IJNSA) Vol.12, No.6, November 2020

$$\dot{x}_1 = \alpha[x_2 - h(x_1)] \qquad (1)$$

$$\dot{x}_2 = x_1 - x_2 + x_3 \qquad (2)$$

$$\dot{x}_3 = -\beta(x_2) \qquad (3)$$

Where: $h(x_1) = bx_1 + \frac{1}{2}(a-b)(|x_1 + 1| - |x_1 - 1|)$ \qquad (4)

For α=10, β=14.87, a=-1.27, and b=-0.1 Chua's exhibits a chaotic attractor.

The Math lap simulation of the Chua Images ciphering system is shown in Figure. 2

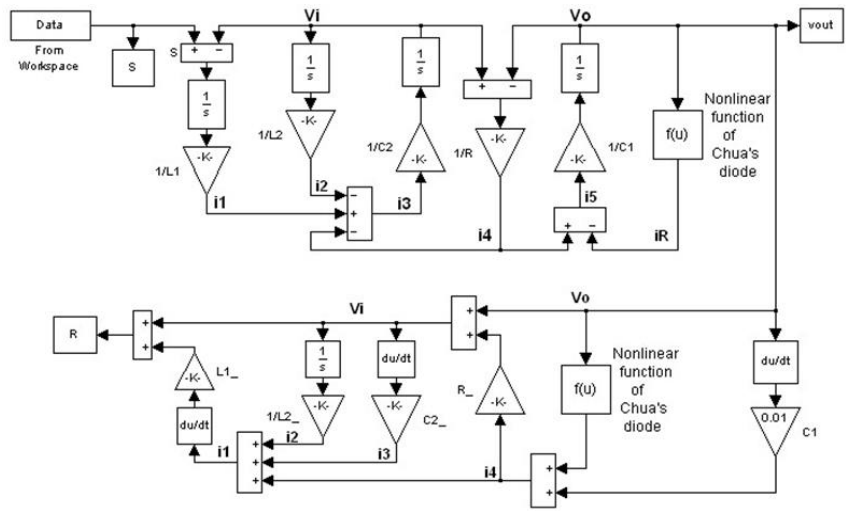

Figure 2: Simulink simulation of Images security based on Chau PRNG

Figure 3 shows the Histogram of Images data (a) and Encrypted Images data (b)

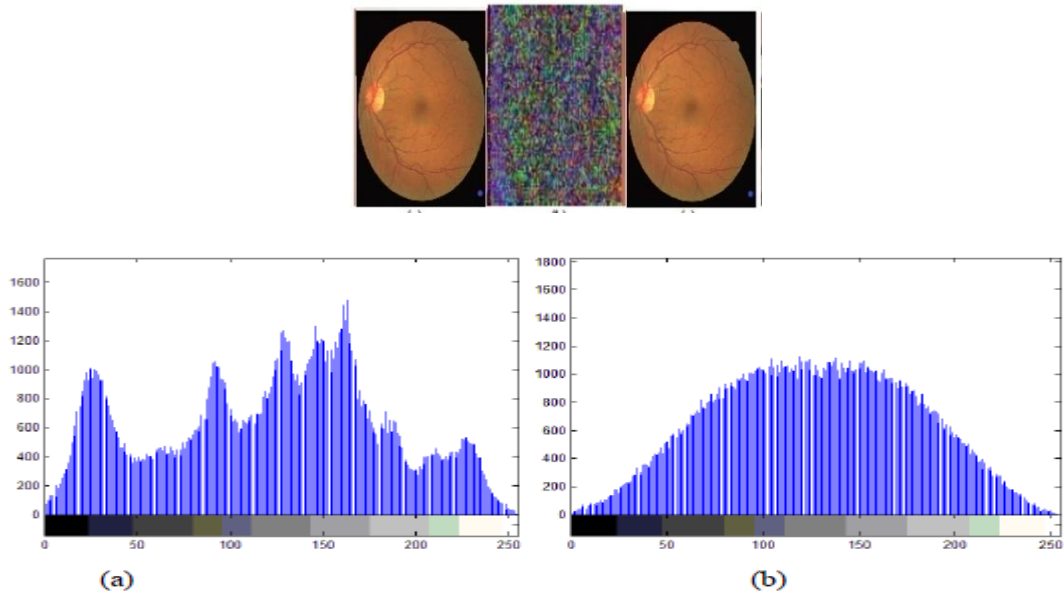

Figure 3: Histogram of Images data (a) and Encrypted Images data (b)

### b. Lorenz Chaotic Generators

Lorenz set of equations are:

$$\dot{x} = a(y - x) \tag{5}$$

$$\dot{y} = bx - y - xz \tag{6}$$

$$\dot{z} = xy - cz \tag{7}$$

When a=10, c=8/3, and b=28 Lorenz exhibits a chaotic attractor.

The Math lap simulation of the Lorenz Images ciphering system is shown in Figure 4.

Figure 4 shows the Simulink implementation of the Lorenz chaotic-based Images cipher system.

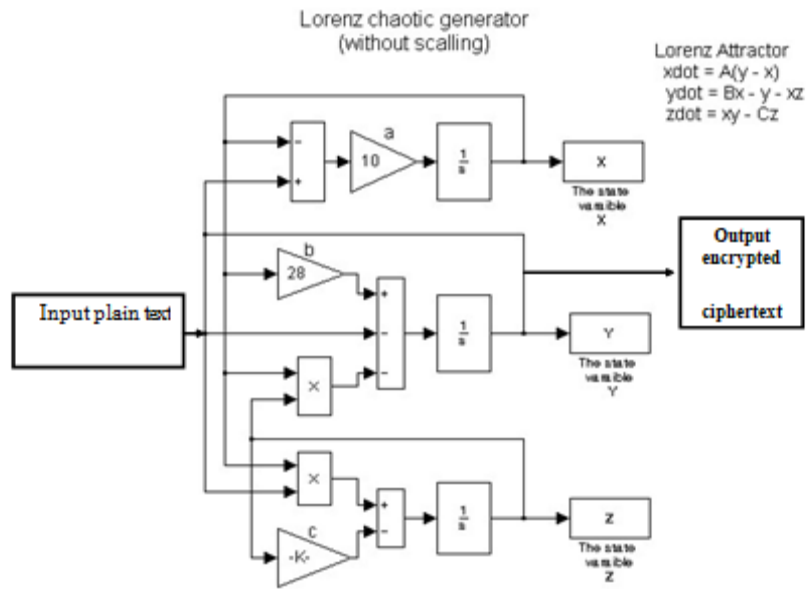

Figure 4: Simulink simulation of Lorenz chaotic generator

### c. Rossler chaotic generators

Rossler system consists following three nonlinear ordinary differential equations:

$$\frac{dx}{dt} = \dot{x}(t) = -y(t) - z(t) \tag{8}$$

$$\frac{dy}{dt} = \dot{y}(t) = x(t) - ay(t) \tag{9}$$

$$\frac{dz}{dt} = \dot{z}(t) = b + z(t)(x(t) - c) \tag{10}$$

Rossler exhibits chaotic attractor with $a = 0.2$, $b = 0.2$, and $c = 5.7$.

The Math lap simulation of the Rossler Images ciphering system is shown in Figure. 5.

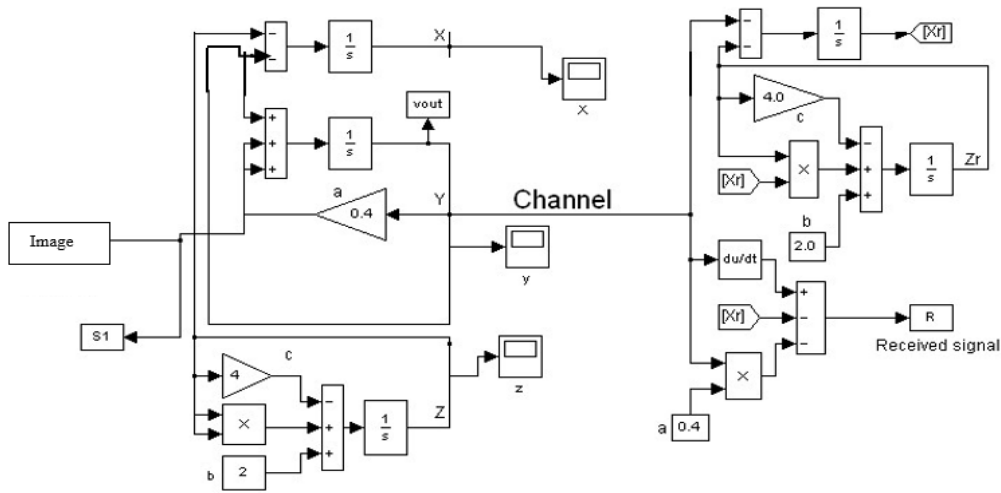

Figure 5: Simulink simulation of Rossler Images security system

### 4.1.2. Discrete Chaotic Generators

**a. Hénon map**

Hénon map is a discrete-time dynamical system whose equations-set are:

$$x(n + 1) = c - ax^2(n) + y(n) \qquad (11)$$

$$y(n + 1) = bx(k) \qquad (12)$$

The map depends on parameters, *a*, *b*, and c, which for values of *a* = 1.4, *b* = 0.3, and *c*=1

The Math lap simulation of the Hénon Images ciphering system is shown in Figure 6

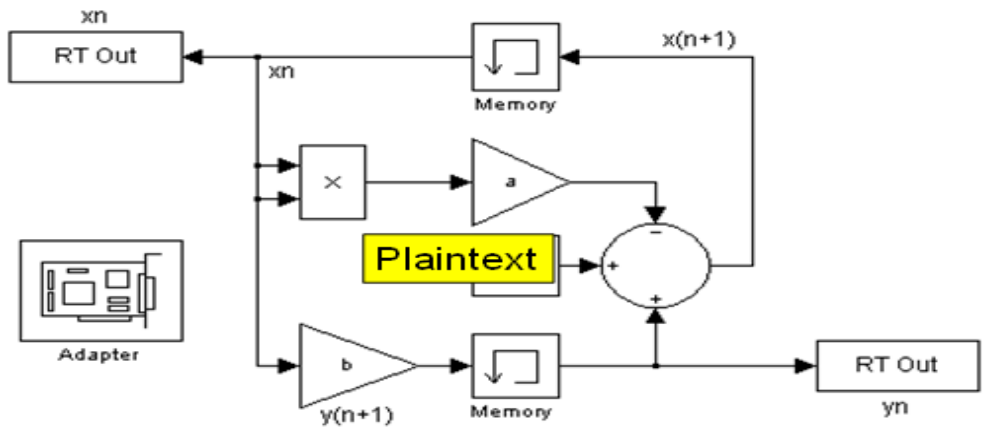

Figure 6: Matlab Simulink Simulation of Images security based on Hénon PRNG

### b. Logistic map

The logistic map is given in the following equation:

$$F: X \rightarrow X, F(x) = \lambda x (1-x) \qquad (13)$$

Where $\lambda$ is the parameter of the logistic map that determines the chaotic behaviour, with $x \in (0, 1)$ and $3.57 \leq \lambda \leq 4$ for good chaotic properties. The value of $\lambda$ parameter is significant because only for the same area of the domain presented above, the logistic map has good chaotic behaviour and this thing can be seen in the bifurcation diagram that helps us to find the good chaotic area. Figure 7. Show the PRNG chaotic behaviour of Logistic algorithm for $3.57 \leq \lambda \leq 4$

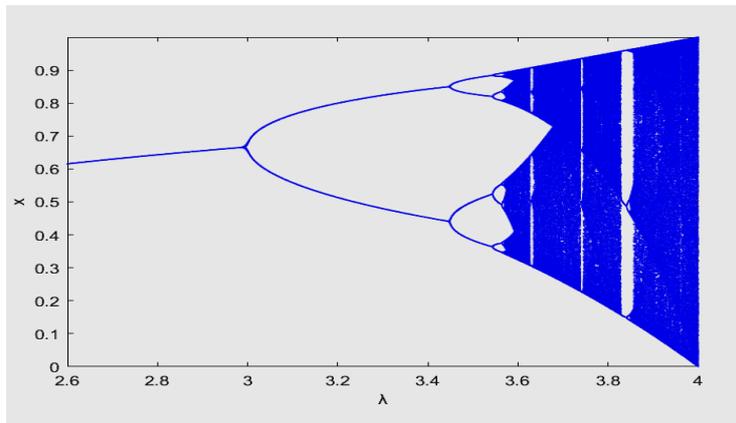

Figure 7: The PRNG chaotic behaviour of Logistic algorithm for $3.57 \leq \lambda \leq 4$

The same PRNG behaviors of the other chaotic algorithms mentioned in this paper are included in the references [15].

Figure. 8 shows the Matlab Simulink Simulation of Logistic map chaotic generators

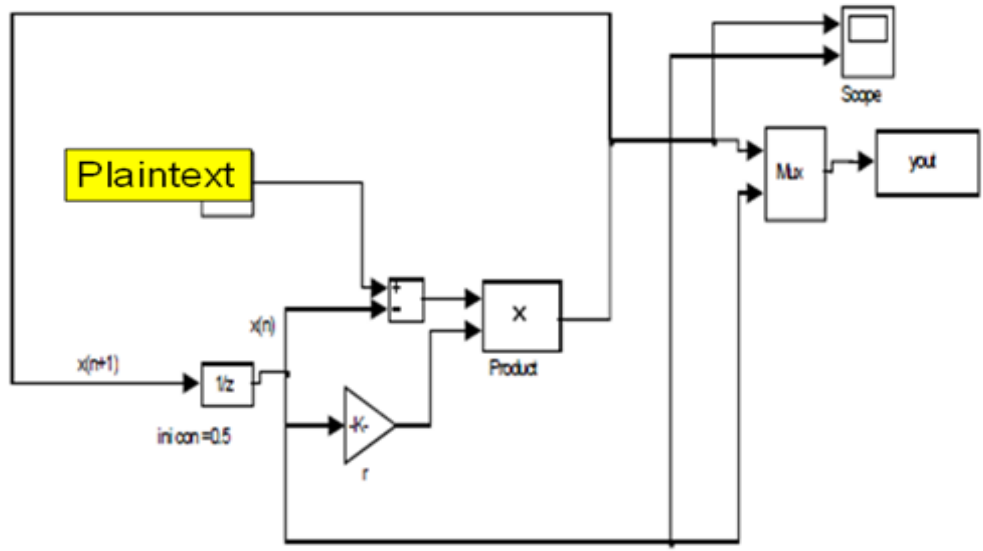

Figure 8: PRNG generation of the Logistic map chaotic generators

## 4.2. Statistical Tests of the Chaotic Generators:

The statistical test gives the probability that the sequence comes from a random number generator. The output of the test is called the probability p-value. If it is determined that the p-value for a test is equal to 1, the sequence tends to have perfect random statistical properties. A p-value equal to zero means that the series is entirely non-random. A significance level ($\alpha=.01$) is chosen for the tests so that if p-value $\geq \alpha$, then the hypothesis that the sequence is random (or pseudo-random) is accepted. If P-value<$\alpha$, then the hypothesis is rejected; i.e., the sequence appears to be non-random.

The NIST tests suite [13] is a statistical package consisting of 16 tests built to test the randomness of binary (arbitrarily long) sequences generated by cryptographic random or pseudo-random bit generators based on hardware or software.

These tests focus on a variety of different types of non-randomness that could exist in a binary sequence. The 16 tests are classified into two categories: (i) non-parameterized tests and (ii) parameterized tests. They assess the presence of a pattern, which, if detected, would indicate that the sequence is non-random. In each test, a p-value is calculated. In all tests in NIST Suite, the significance level $\alpha$ is set to 1%. A p-value of less than $\alpha$ would mean that the sequence is non-random with a confidence of 99%. If a p-value is more significant than $\alpha$, we accept the sequence as random with a confidence of 99%.

Overall NIST tests include: (Block Frequency, Cumulative sums, Fast Fourier Transform, Frequency (Mono-bit), Lempel-Ziv compression, Linear complexity, Longer Runs of 1's, Maurer's Universal Statistics, Non-Overlapping, Approximate Entropy, Overlapping Template Matching, Random Excursions, Ranks, Runs, and Serials).

Among these tests, we implemented the tests in table 1:

Table 1: Implemented NIST PRNG statistical tests

| S | Test | Meaning |
|---|---|---|
| 1 | Frequency mono bits | Existence of too many zeroes or ones. |
| 2 | Block tests | Requires continuously comparing adjacent blocks of bits, and if identical the 2nd block is tested |
| 3 | Serial tests | Determine whether the number of occurrences of m-bits overlapping patterns is approximately the same as would be expected for a random sequence |
| 4 | Runs Large (small) | The total number of runs indicates that the bitstream oscillation is too rapid (too slow). |
| 5 | Rank | Deviation of the rank distribution from a corresponding random sequence, due to periodicity. |
| 6 | Longest Runs Of Ones | Deviation of the distribution of long runs of ones. |
| 7 | Random Excursions | Deviation from the distribution of the number of visits of a random walk to a particular state. |
| 8 | Random Excursion Variant | Deviation from the distribution of the total number of visits (across many random walks) to a certain state. |
| 9 | Cumulative Sums | Existence of too many zeroes/ones at the sequence beginning. |
| 10 | Linear Complexity | Deviation from the distribution of the linear complexity for finite length (sub) strings. |

International Journal of Network Security & Its Applications (IJNSA) Vol.12, No.6, November 2020

Critical parameters and initial condition values for each tested chaotic algorithm are discussed. These critical parameters are selected for each model. All discussed chaotic generator models are capable of securing Images and all other types of information.

The 5-previous chaotic PRNGs have been implemented using Matlab .m files and Simulink. The PRNGs outputs are in decimal values.

After $N \geq 106$ iterations, and in order to construct a binary bit sequence, the mean of each PRNG y is calculated and the output is converted to '0' and '1' using the formula (15):

$B_n = 1$ if output is $x_n > y$, and $B_n = 0$, otherwise, for $n = (1, 2 ... N)$  (14)

Finally, a bit sequence $B_n = \{b_1, b_2, b_3..., b_N\}$ is produced. This sequence is statistically tested.

The output of a single chaotic algorithm passed a limited number of the statistical tests where the value of $p < 1$, as shown in table 2.

Table 2: The statistical tests of single chaotic algorithms

| Single Chaotic Algorithm | Frequency (p value) | Block tests (p-value) | Runs test (p value) | Serial tests (p value) |
|---|---|---|---|---|
| Chau | .1581 | .9891 | .1518 | .1948 |
| Lorenz | .6626 | .9891 | .7071 | .4549 |
| Rossler | .3626 | .5851 | .3707 | .1298 |
| Henon | .18 | .1111 | .2759 | .1145 |
| Logistics | .6488 | .6805 | .6293 | .1948 |

## 5. IMAGES ENCRYPTION USING MIXED / HYBRID CRYPTOGRAPHY ALGORITHMS

An effective Image encryption can be implemented using a mix of cryptography algorithms. Two ways can implement this mix:

1) Mix Two Chaotic algorithms (e.g. Logistic, Lorenz, Rossler, Henon, or Chau maps) as in [15] [17].
2) The mix of traditional cryptography (e.g. DES, 3DES, or AES) and chaotic algorithms (e.g. Logistic, Lorenz, Rossler, Henon, or Chau maps).

A block diagram of this mix in shown in Figure 9.

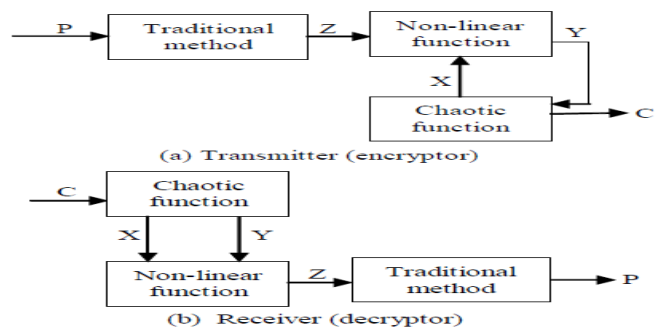

Figure 9: Hybrid Images security traditional and chaotic cypher system

According to this scheme at the transmitter side, the Images plaintext *P* is encrypted by an encryption rule, which uses a traditional method to produce signal *Z*, non-linear function to produce signal *Y*, and the state produced by the chaotic system to produce cipher-text *C* in the transmitter.

The scrambled output signal is used further to drive the chaotic system such that the chaotic dynamics is changed continuously in a very complicated way. Then another state variable of the chaotic system in the transmitter is transmitted through the channel.

At the receiver side, the reconstruction of the plaintext is done by decrypting the input by using the reverse of encryption method.

### 5.1. Mixed Chaotic Algorithms using Logistics Maps

We used Matlab to simulate Image security using mix 2-Logistics algorithms whose binary PRNG sequences are tested in section 4. The Images plaintext is converted to binary data and *XOR*ed to the PRNG output, and the results are tested.

Figure 10 shows the block diagram of the Images security using two Logistic maps PRNG, where the output from the mixed PRNG is *XOR*ed with the plaintext (Images) to produce the cipher-text.

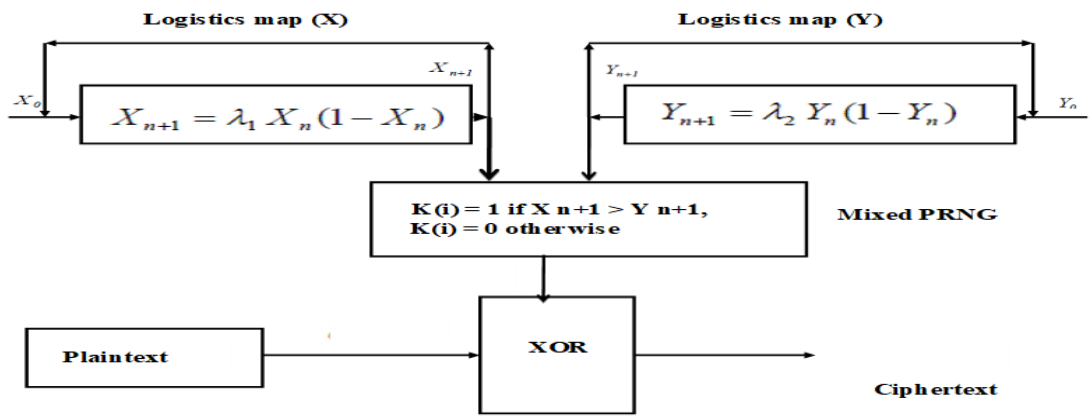

Figure 10: Block diagram of the Images security using two Logistic maps PRNG.

### 5.2. Mixed Chaotic Algorithms Using Henon Maps

The proposed PRNG based on two chaotic Henon maps, starting from random independent initial conditions ($X_n, Y_n \in (0, 1)$ and $X_n \neq Y_n$) where

$$x_1(k+1) = a_1 - b_1 z(k) + z(k) \tag{16}$$

$$y_1(k+1) = a_2 - b_2 h(k) + h(k) \tag{17}$$

The binary bit sequence $G_n$, n= (1, 2…N), (N=$10^5$-$10^6$) is generated by comparing the outputs of both the logistic maps as follows: $G_n$= 1 if $X_n > Y_n$, $G_n$=0 otherwise.

The set of initial conditions ($X_n, Y_n \in (0, 1)$, and $X_n \neq Y_n$) serves as the seed for the PRNG. Figure 11 shows the schematic block diagram of the proposed PRNG.

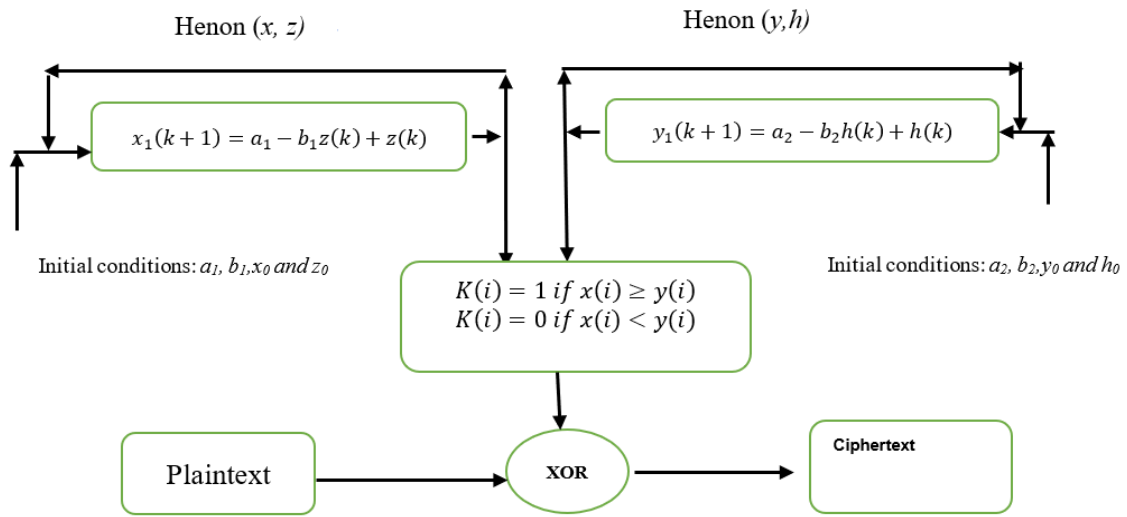

Figure 11: Schematic diagram of the PRNG with two PRNG Henon chaotic maps.

The output of hybrid mixed chaotic algorithm passed most of the statistical tests where the value of p< 1 as shown in table 3.

Table 3: Implemented NIST PRNG statistical tests

| S | Test | Mixed Henon Map | Mixed Logistic Map |
|---|---|---|---|
| 1 | Frequency mono bits | .8594 | .6863 |
| 2 | Block tests | .121 | .9793 |
| 3 | Serial tests | .5162 | .5058 |
| 4 | Runs Large (small) | .4521 | .5561 |
| 5 | Rank | .5094 | .0556 |
| 6 | Longest Runs Of Ones | .4121 | .1509 |
| 7 | Random Excursions | .5161 | .2121 |
| 8 | Random Excursion Variant | .5021 | .3516 |
| 9 | Cumulative Sums | .4523 | .6021 |
| 10 | Linear Complexity | .2345 | .5643 |

## 6. CONCLUSIONS

In this paper, image security, PRNG concepts, and characterization using statistical and frequency test-suits, chaos theory and its application for Image security are provided, and an approach for Image encryption and decryption based on different chaos algorithms is discussed.
There are several options available for simulation and analyzing the randomness of the most famous chaotic algorithms and pseudo-random bit generators. In this paper, Matlab (.m files and Simulink) are used for the simulation, realization, and implementation of chaotic algorithms, and the NIST 16 statistical test suit is used for the statistical tests.

Several statistical tests designed to detect the specific characteristics needed for truly random sequences should be subjected to cipher-text in order to gain confidence that the use of image protection (Lorenz, Chau, and Roosler) and discrete (Henon and logistic) chaotic cipher machine models continues;

We statistically tested the PRNG and applied them for Image encryption. All keys and cipher-text outputs are tested using the NIST test suit. It was shown that they passed a limited number of tests.

We have shown how to PRNG using: single continuous, single discreet, and hybrid mixed chaotic algorithms. We conclude that the randomness properties of Single Chaotic algorithms one dimension 1D are not acceptable. They passed only 4 NIST tests and Failed in the remaining 12 tests.

The research proposed mixed/hybrid PRNG based on 2 Chaotic algorithms (Logistics and Henon) operating with different initial conditions and encryption keys. All keys and cipher-text outputs are tested using the NIST test suit. The hybrid algorithm passed 10 of NIST tests.

It was shown that hybrid mixed are 'computationally secure" than single chaotic algorithms, i.e. secure because the time and/ or cost of defeating the security are too high to be feasible.

In future researches, we can simulate new applications for chaotic algorithms such as biometrics authentication, hash functions for a digital signature that work better than the current existing hash functions. New hash functions based on Chaotic Maps can be simulated and tested.

Utilize Chaotic Maps to design new cryptography algorithms depending on higher-order dynamic equations and mix/hybrid between chaotic algorithms as well as a mix between classic (DES, 3DES, AES…) and chaotic algorithms (Hua, Lorenz, Rossler, Logistics, Henon,)

## ACKNOWLEDGEMENTS

The writer would like to thank Prof.WafaeBoghdady and express her appreciation and gratitude for his scientific assistance, his consistent encouragement, and support

## AUTHOR

**Seham Ebrahim** Obtained her MSc degree f from Ain-Shams UniversityEgypt. And Ph.D. from Cairo University. Currently, work asa lecturer in Modern Academy for Engineering and Information Technology. Interested in Network Security, Machine Learning, and Computer Technology.

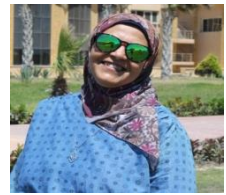